\documentclass[final,5p,times,twocolumn]{elsarticle}
\usepackage{amsmath}

\usepackage{graphicx}
\usepackage{dcolumn}
\usepackage{bm}
\def\jpsi{{J/\psi}}
\def\jpsio{{J/\psi\bigl[\bigl.^1\hspace{-1mm}S^{(8)}_0,\bigl.^3\hspace{-1mm}S^{(8)}_1\bigr]}}

\def\be{\begin{equation}}
\def\ee{\end{equation}}
\def\bea{\begin{eqnarray}}
\def\eea{\end{eqnarray}}

\def\a{\alpha}

\begin{document}
\begin{frontmatter}
\title{QCD corrections to ${J/\psi}$ production via color-octet states at
the Tevatron and LHC}
\author{Bin Gong$^{1}$, Xue Qian Li$^2$ and Jian-Xiong Wang$^{1,3}$}
\address{$^1$Institute of High Energy Physics, Chinese Academy of Sciences, P.O. Box
918(4), Beijing, 100049, China.\\
$^2$Department of Physics, Nankai University, Tianjin 300071, China.\\
$^3$Theoretical Physics Center for Science Facilities, CAS, Beijing, 100049,
China. }
\date{\today}

\begin{abstract}
The Next-To-Leading-Order (NLO) QCD corrections to $\jpsi$
production via S-wave color-octet states at the Tevatron and LHC are
calculated. There are only slight changes to the transverse momentum $p_t$ distributions of $\jpsi$ production
and polarization. By fitting the $p_t$ distribution of $\jpsi$ production measured at Tevatron
with both color-singlet and color-octet included, we obtain the color-octet matrix
elements at NLO. The resulting $p_t$ distribution of $\jpsi$ polarization still
does not coincide with the experimental data. Therefore, we have reached
the conclusion that the large discrepancy of $\jpsi$ polarization
between theory and experimental data cannot be solved by just including NLO
corrections within non-relativistic QCD framework and then
one has to explore solution from different angles.
\end{abstract}
\end{frontmatter}

%

Since its discovery in 1974, study on $J/\psi$ production never ends,
it seems that this field is still not fully understood. To solve the
large discrepancy between the experimental data and theoretical predictions
on the $p_t$ distribution of $J/\psi$ production at Tevatron, a color-octet
mechanism \cite{Braaten:1994vv} which increases the $p_t$
distribution was proposed based on the non-relativistic QCD (NRQCD) \cite%
{Bodwin:1994jh}. However, despite the developments and successes of
NRQCD, its predictions are not always satisfactory, namely fail to
give right values for some processes. The authors of
Ref.~\cite{Klasen:2001cu} find that the DELPHI \cite{deBoer:2003xm}
data for $J/\psi$ production in $\gamma \gamma \rightarrow J/\psi X$
evidently favor the NRQCD formalism as long as the color-octet
contributions are included. By contrast, an analysis on charmonium
production at fixed-target experiments was performed with NRQCD up
to NLO \cite{Maltoni:2006yp}, and it is indicated that the fraction
of color-octet which is needed to describe the data is only about
$1/10$ of that required for explaining the Tevatron experiment.
Whereas, the experimental results on inelastic $J/\psi$
photoproduction at the DESY ep collider HERA are adequately
described by the color-singlet mechanism alone
once higher-order QCD corrections are included \cite{Ko:1996xw,kramer:1995nb}%
. At B-factories, ${J/\psi\bigl[^3\hspace{-1mm}P^{(8)}_J\bigr]}$
production suggested by the authors of Ref.~\cite{Braaten:1995ez}
has not been observed. Even more seriously, the LO NRQCD calculation
predicts a sizable transverse polarization rate for large $p_t$
${J/\psi}$ \cite{beneke:96yr} whereas the Tevatron measurement at
Fermilab \cite{Abulencia:2007us} displays a slight longitudinal
polarization at large $p_t$.

On the other hand, obvious discrepancies between LO predictions \cite%
{Braaten:2002fi, Liu:2002wq} and experimental results ~\cite%
{Abe:2002rb,Aubert:2005tj} for single and double charmonia
productions at B-factories drew close attention of theorists.
Further studies indicate that they may be resolved by including
higher order corrections: both NLO QCD and relativistic corrections
\cite{Braaten:2002fi, Zhang:2005ch, Gong:2008ce}, or at least the
trouble is alleviated. Recently, the NLO QCD corrections to
${J/\psi}$ hadronproduction are calculated in
Refs.~\cite{Campbell:2007ws,Gong:2008sn} and the results show that
the production rate of ${J/\psi}$ at the larger transverse momentum
$p_t $ region is much increased. The NLO process $g g\rightarrow
{J/\psi} c \overline{c}$ is calculated in
Refs.~\cite{Hagiwara:2007bq,Artoisenet:2007xi} and it causes a
sizable contribution to the $p_t$ distribution.
The NLO QCD corrections to ${J/\psi}$ polarization via color-singlet
at Tevatron and LHC have been calculated in Ref.~\cite{Gong:2008sn}
and the results show that the ${J/\psi}$ polarization status
drastically changes from transverse-polarization dominance at LO
into longitudinal-polarization dominance at NLO. However, such
changes still cannot explain the data yet. Since the NLO corrections
are so important for the $p_t$ distribution and polarization status,
when one considers contributions from the color-octet mechanism to
explain data, it is obviously necessary to investigate whether NLO
corrections can seriously affect ${J/\psi}$ production in that case.
In this letter, we calculate the NLO QCD corrections to the
${J/\psi}
$ production via color-octet states ${J/\psi\bigl[\bigl.^1\hspace{-1mm}%
S^{(8)}_0,\bigl.^3\hspace{-1mm}S^{(8)}_1\bigr]}$ at Tevatron and LHC where
the Feynman Diagram Calculation package (FDC) \cite{FDC} is employed.
However, it is noted that ${J/\psi\bigl[^3\hspace{-1mm}P^{(8)}_J\bigr]}$ is
not included since the part for dealing with P-wave loop processes in FDC is
not completed yet.

According to the NRQCD factorization formalism, the inclusive cross section
for direct ${J/\psi}$ production in hadron-hadron collision is expressed as
\begin{eqnarray}
\sigma[pp\rightarrow {J/\psi}+X]&=\sum\limits_{i,j,n}\int \mathrm{d}x_1%
\mathrm{d}x_2 G_{i/p}G_{j/p}  \nonumber \\
&\times\hat{\sigma}[i+j\rightarrow (c\bar{c})_n +X]{\langle\mathcal{O}%
^H_n\rangle},
\end{eqnarray}
where $p$ is either a proton or an antiproton, the indices $i, j$
run over all partonic species and $n$ denotes the color, spin and
angular momentum states of the intermediate $c\bar{c}$ pair. The
short-distance contribution $\hat{\sigma}$ can be perturbatively
calculated order by order in $\alpha_s$. The hadronic matrix
elements ${\langle\mathcal{O}^H_n\rangle}$ are related to the
hadronization probabilities from the state $(c\bar{c})_n$ into
${J/\psi}$ which are fully governed by the non-perturbative QCD
effects. In the following, $\hat{\sigma}$ represents the
corresponding partonic cross section.

At LO, there are three partonic processes:

\begin{align}
&L1:g g \rightarrow \jpsio g,
&L2:g q \rightarrow \jpsio q, \nonumber \\
&L3:q \overline{q} \rightarrow \jpsio g. & \nonumber
\end{align}
where $q$ runs over all possible light quarks or anti-quarks: $%
u,d,s,\overline{u},\overline{d},\overline{s}$.

The NLO corrections include virtual and real corrections. There
exist UV, IR and Coulomb singularities in the calculation of the
virtual corrections. The UV-divergences from self-energy and
triangle diagrams are removed by the
renormalization procedure. Here we adopt the dimensional renormalization scheme and technique used in Ref.~\cite%
{Petrelli:1997ge} without performing an explicit matching between the cross sections calculated in 
perturbative QCD and perturbative NRQCD.
The renormalization constants $Z_{m}$, $Z_{2}$, $Z_{2l}$
and $Z_{3}$ which correspond to charm quark mass $m_{c}$, charm-field $\psi
_{c}$, light quark field $\psi _{q}$ and gluon field $A_{\mu }^{a}$ are
defined in the on-mass-shell(OS) scheme while $Z_{g}$ for the QCD gauge
coupling $\alpha _{s}$ is defined in the modified-minimal-subtraction($%
\overline{\mathrm{MS}}$) scheme:
\begin{eqnarray}
\delta Z_{m}^{OS} &=&-3C_{F}\displaystyle\frac{\alpha _{s}}{4\pi }\left[ %
\displaystyle\frac{1}{\epsilon _{UV}}-\gamma _{E}+\ln \displaystyle\frac{%
4\pi \mu _{r}^{2}}{m_{c}^{2}}+\frac{4}{3}\right] ,  \nonumber \\
\delta Z_{2}^{OS} &=&-C_{F}\displaystyle\frac{\alpha _{s}}{4\pi }\left[ %
\displaystyle\frac{1}{\epsilon _{UV}}+\displaystyle\frac{2}{\epsilon _{IR}}%
-3\gamma _{E}+3\ln \displaystyle\frac{4\pi \mu _{r}^{2}}{m_{c}^{2}}+4\right]
,  \nonumber \\
\delta Z_{2l}^{OS} &=&-C_{F}\displaystyle\frac{\alpha _{s}}{4\pi }\left[ %
\displaystyle\frac{1}{\epsilon _{UV}}-\displaystyle\frac{1}{\epsilon _{IR}}%
\right] ,  \nonumber \\
\delta Z_{3}^{OS} &=&\displaystyle\frac{\alpha _{s}}{4\pi }\left[ (\beta
_{0}-2C_{A})\left( \displaystyle\frac{1}{\epsilon _{UV}}-\displaystyle\frac{1%
}{\epsilon _{IR}}\right) \right] , \\
\delta Z_{g}^{\overline{\mathrm{MS}}} &=&-\displaystyle\frac{\beta _{0}}{2}%
\displaystyle\frac{\alpha _{s}}{4\pi }\left[ \displaystyle\frac{1}{\epsilon
_{UV}}-\gamma _{E}+\ln (4\pi )\right] ,  \nonumber
\end{eqnarray}%
where $\gamma _{E}$ is the Euler constant, $\beta _{0}=\frac{11}{3}C_{A}-%
\frac{4}{3}T_{F}n_{f}$ is the one-loop coefficient of the QCD beta function
and $n_{f}=3$ is the number of active quark flavors.
$\mu _{r}$ is the renormalization scale.


There are 267 (for the ${\bigl.^1\hspace{-1mm}S^{(8)}_0}$ state) and 413 (for the ${%
\bigl.^3\hspace{-1mm}S^{(8)}_1}$ state) NLO diagrams for process (L1),
including counter-term diagrams, while for both processes (L2)
and (L3), there are 49 (for the ${\bigl.^1%
\hspace{-1mm}S^{(8)}_0}$ state) and 111 (for the ${\bigl.^3\hspace{-1mm}%
S^{(8)}_1}$ state) NLO diagrams altogether. Diagrams where a virtual gluon
line connects the quark pair possess a Coulomb singularity, which can be
isolated and attributed into the renormalization of the $c\bar{c}$ wave
function.

For each process, by summing over contributions from all diagrams, the
virtual corrections to the differential cross section can be expressed as
\begin{equation}
\displaystyle\frac{\mathrm{d}\hat{\sigma}^{V}_i}{\mathrm{d}t} \propto 2%
\mathrm{Re}(M^B_iM^{V*}_i),
\end{equation}
where $M^B_i$ is the amplitude of process $(i)$ at LO, and $M^V_i$ is the
renormalized amplitude of corresponding process at NLO. $M^V_i$ is UV and
Coulomb finite, but still contains IR divergences.

It is noteworthy that to obtain a full cancelation of
IR-singularities in the calculation, the three sub-processes (L1),
(L2) and (L3) tangle together and must be considered simultaneously.
In addition, there are eight tree processes involved in the real
corrections:
\begin{eqnarray}
gg\rightarrow {J/\psi }\left[ {^{1}\hspace{-1mm}S_{0}^{(8)},^{3}\hspace{-1mm}%
S_{1}^{(8)}}\right] gg,~~gg &\rightarrow &{J/\psi }\left[ {^{1}\hspace{-1mm}%
S_{0}^{(8)},^{3}\hspace{-1mm}S_{1}^{(8)}}\right] q\overline{q},  \nonumber \\
gq\rightarrow {J/\psi }\left[ {^{1}\hspace{-1mm}S_{0}^{(8)},^{3}\hspace{-1mm}%
S_{1}^{(8)}}\right] gq,~~q\overline{q} &\rightarrow &{J/\psi }\left[ {^{1}%
\hspace{-1mm}S_{0}^{(8)},^{3}\hspace{-1mm}S_{1}^{(8)}}\right] gg,  \nonumber
\\
q\overline{q}\rightarrow {J/}\psi \left[ {^{1}\hspace{-1mm}S_{0}^{(8)},^{3}%
\hspace{-1mm}S_{1}^{(8)}}\right] q\overline{q},~~q\overline{q} &\rightarrow &%
{J/\psi }\left[ {^{1}\hspace{-1mm}S_{0}^{(8)},^{3}\hspace{-1mm}S_{1}^{(8)}}%
\right] q^{\prime }\overline{q}^{\prime },  \nonumber \\
qq\rightarrow {J/\psi }\left[ {^{1}\hspace{-1mm}S_{0}^{(8)},^{3}\hspace{-1mm}%
S_{1}^{(8)}}\right] qq,~~qq^{\prime } &\rightarrow &{J/\psi }\left[ {^{1}%
\hspace{-1mm}S_{0}^{(8)},^{3}\hspace{-1mm}S_{1}^{(8)}}\right] qq^{\prime },
\nonumber
\end{eqnarray}%
where $q,q^{\prime }$ denote light quarks (anti-quarks) with different
flavors. Phase space integrations of above processes generate IR
singularities, which are either soft or collinear and can be conveniently
isolated by slicing the phase space into different regions. Here we adopt
the two-cutoff phase space slicing method \cite{Harris:2001sx} to deal with
the problem. 
Then the real cross section can be written as
\begin{equation}
\sigma ^{R}=\sigma ^{S}+\sigma ^{HC}+\sigma ^{H\overline{C}}+\sigma
_{add}^{HC}.
\end{equation}%
It is observed that the IR singularities from one real process may be
factorized into different parts and each of them should be added into the
cross sections of different LO processes. This is the reason why we have to
calculate the NLO corrections to the three LO processes together.

$\hat{\sigma}^S$ from the soft region contains soft singularities
and is calculated analytically under the soft approximation. One
should notice that, unlike color-singlet case, the soft
singularities caused by emitting a soft gluon from the charm quark
pair in the S-wave color-octet exist and the factorized matrix
element is the same as the case where a soft gluon is emitted from a
gluon. $\sigma^{HC}$ from the hard collinear region contains
collinear singularities which are factorized out and the
singularities are partly absorbed into redefinition of the parton
distribution function (PDF) (usually called as mass factorization
\cite{Altarelli:1979ub}). Here we adopt the scale dependent PDF
using the $\overline{\mathrm{MS}}$ convention given in Ref.
\cite{Harris:2001sx}. After redefining the PDF, an additional finite
term $\sigma^{HC}_{add}$ is separated out. The hard non-collinear
part $\sigma^{H\overline{C}}$ is IR finite. Finally, all the IR
singularities are canceled and
$\hat{\sigma}^S+\hat{\sigma}^{HC}+\hat{\sigma}^{V}$ is IR finite.

To obtain the transverse momentum $p_t$ distribution of ${J/\psi}$, a
transformation of integration variables ($\mathrm{d} x_2 \mathrm{d} t
\rightarrow J\mathrm{d} p_t \mathrm{d} y$) is needed. Then we have
\begin{eqnarray}
\displaystyle\frac{\mathrm{d} \sigma}{\mathrm{d} p_t}= \sum \int J \mathrm{d}
x_1 \mathrm{d} y G_{\alpha}(x_1,\mu_f)G_{\beta}(x_2,\mu_f) \displaystyle%
\frac{\mathrm{d} \hat \sigma}{\mathrm{d} t},
\end{eqnarray}
where $y$ is the rapidity of ${J/\psi}$ in the laboratory frame and $\mu_f$
is the factorization scale. The polarization parameter $\alpha$ is defined
as:
\begin{equation}
\alpha(p_t)=\frac{{\mathrm{d}\sigma_T}/{\mathrm{d} p_t}-2 {\mathrm{d}\sigma_L%
}/{\mathrm{d} p_t}} {{\mathrm{d}\sigma_T}/{\mathrm{d} p_t}+2 {\mathrm{d}%
\sigma_L}/{\mathrm{d} p_t}}.
\end{equation}
To evaluate $\alpha(p_t)$, the polarization of ${J/\psi}$ must be explicitly
retained in the calculation. The partonic differential cross section for a
polarized ${J/\psi}$ is expressed as:
\begin{equation}
\displaystyle\frac{\mathrm{d} \hat{\sigma}_{\lambda}}{\mathrm{d} t}=
a~\epsilon(\lambda) \cdot \epsilon^*(\lambda) + \sum_{i,j=1,2} a_{ij} ~p_i
\cdot \epsilon(\lambda) ~p_j \cdot \epsilon^*(\lambda),  \label{eqn:polar}
\end{equation}
where $\lambda=T_1,T_2,L$.
$\epsilon(T_1),~\epsilon(T_2),~\epsilon(L)$ are the two transverse
and longitudinal polarization vectors of ${J/\psi}$ respectively,
and the polarizations of all the other particles are summed over in
n-dimension. One can find that $a$ and $a_{ij}$ are finite when the
virtual corrections and real corrections are properly handled as
aforementioned. The gauge invariance is explicitly checked that the
amplitude is exactly zero as the gluon polarization vector being
replaced by its 4-momentum in the final numerical calculation.

\begin{figure}[tbp]
\center{
\includegraphics*[scale=0.45]{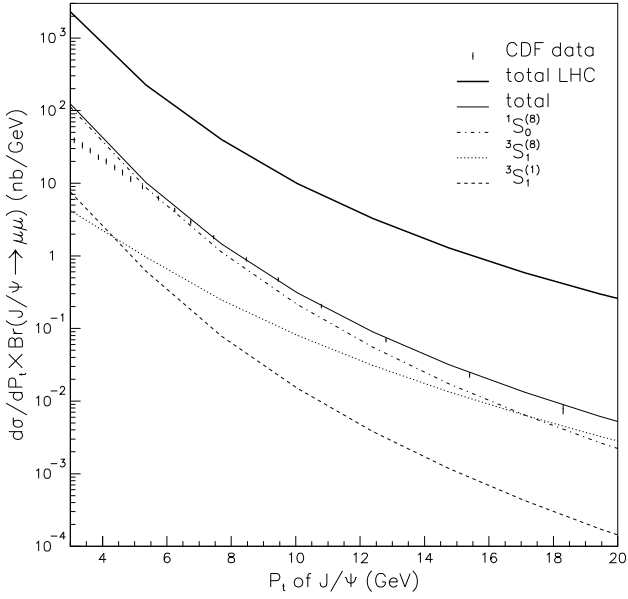}\caption {\label{fig:fit}Transverse momentum
distribution of prompt $\jpsi$ production at Tevatron.
CDF data is from ref~\cite{Abulencia:2005us}.  The
center-of-mass energy are 1.96 TeV at Tevatron and 14 TeV at LHC.}}
\end{figure}
\begin{figure}[tbp]
\center{
\includegraphics*[scale=0.45]{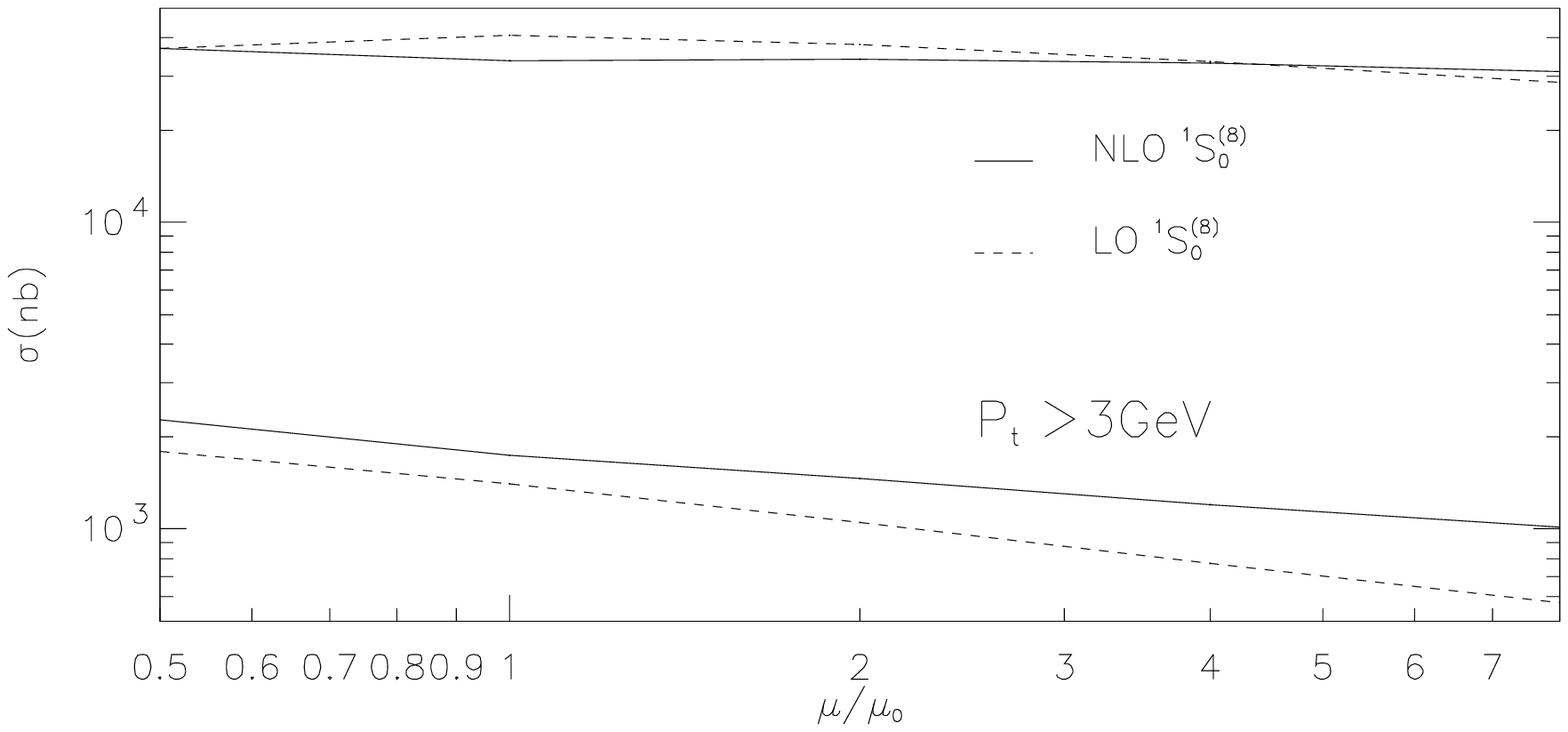}\\
\includegraphics*[scale=0.45]{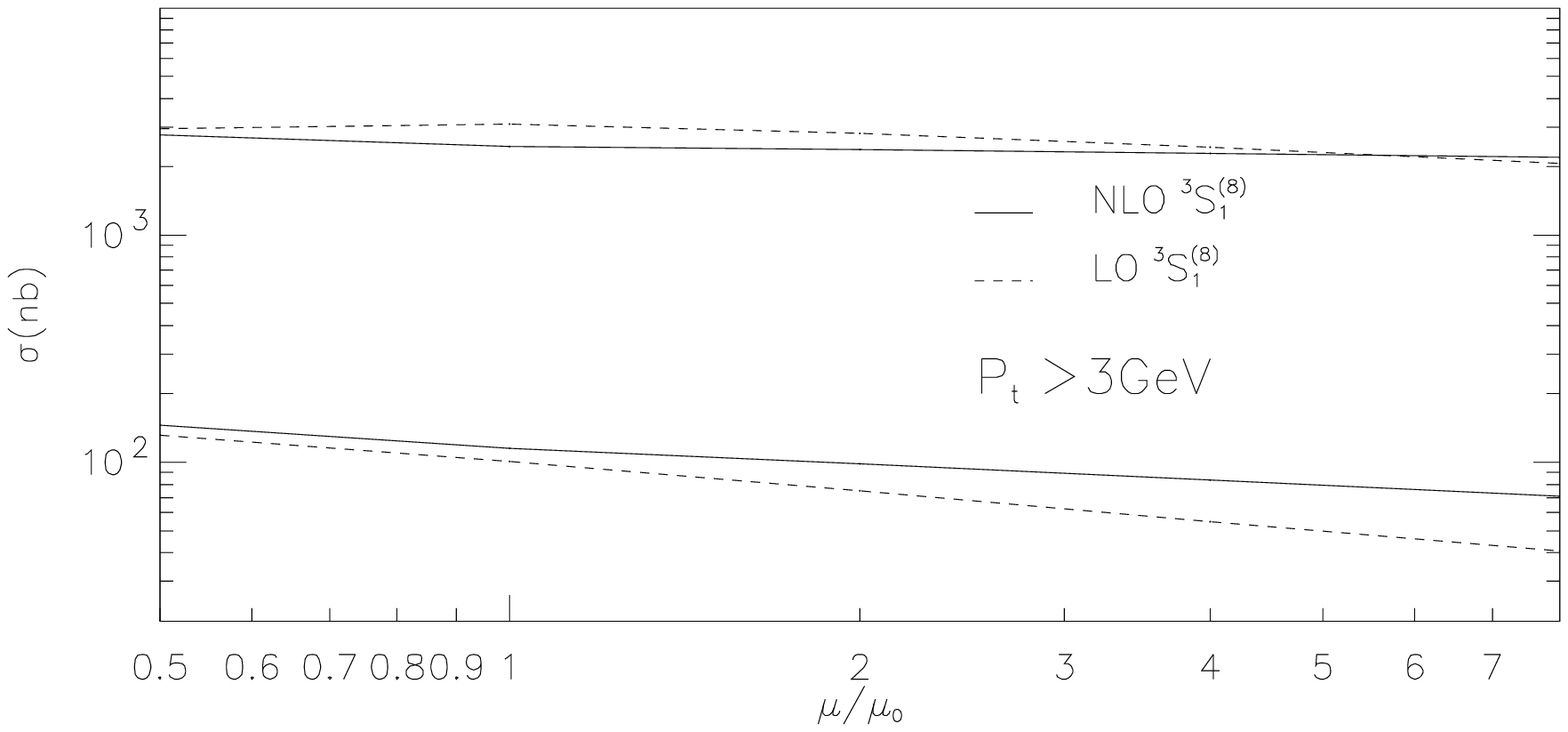}\caption {\label{fig:total}
The partial cross section (with cut conditions) of $\jpsi$ hadronproduction at LHC (upper
curves) and Tevatron (lower curves), 
as a function of $\mu$ with
$\mu_r=\mu_f=\mu$ and $\mu_0=\sqrt{(2m_c)^2+p_t^2}$.}}
\end{figure}
\begin{figure}[tbp]
\center{
\includegraphics*[scale=0.45]{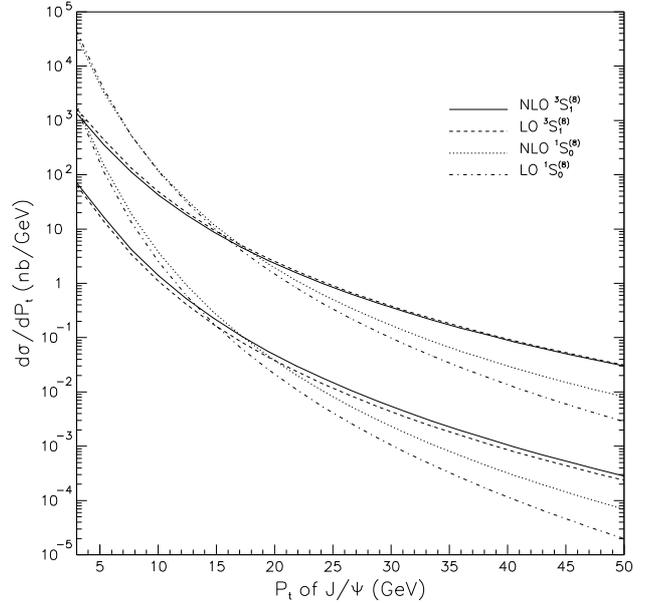}\caption {\label{fig:pt} Transverse
momentum distribution of $\jpsi$ production
with $\mu_r=\mu_f=\mu_0$ at LHC (upper curves)
and Tevatron (lower curves).}}
\end{figure}

In our numerical computations, the CTEQ6L1 and CTEQ6M PDFs \cite{cteq}
, and the corresponding fitted value $\a_s(M_Z)=0.130$ and $\a_s(M_Z)=0.118$
are used for LO and NLO calculations respectively. The charm quark mass is
set as $1.5 \mathrm{~GeV}$. The two phase space cutoffs $\delta_s=10^{-3}$ and $%
\delta_c=\delta_s/50$ are chosen, and the invariance for different values of
$\delta_s$ and $\delta_c$ is obviously observed within the error tolerance.
All the results in this paper are restricted to the NRQCD applicable domain $p_t>3$ GeV, and $%
|y_{{J/\psi}}|<3$ for LHC, $|y_{{J/\psi}}|<0.6$ for Tevatron respectively.

By fitting the $p_t$ distribution of prompt ${%
J/\psi}$ production measured at Tevatron ~\cite{Abulencia:2005us}, the NRQCD
matrix elements ${\langle\mathcal{O}^H_n\rangle}$ are determined as ${\langle%
\mathcal{O}^\psi_8(\bigl.^3\hspace{-1mm}S_1)\rangle}=0.0021 \mathrm{~GeV}^3$
and ${\langle\mathcal{O}^\psi_8(\bigl.^1\hspace{-1mm}S_0)\rangle}=0.075
\mathrm{~GeV}^3$, and the results are shown in Fig.~\ref{fig:fit}. In the
fitting procedure, contributions from both color singlet at NLO\cite{Gong:2008sn} and octet states
at NLO are included. However, it is worth noticing that we have to abandon the
experimental data with $p_t<6$ GeV, since it is impossible to obtain a
satisfactory $p_t$ distribution in terms of a unique ${\langle\mathcal{O}%
^H_n\rangle}$ value. In addition, one should consider an additional
contribution of the feed-down from $\psi^\prime$ which may bring up an extra
factor $B(\psi^\prime\rightarrow {J/\psi}+X)\times{\langle\mathcal{O}%
^{\psi^\prime}_n\rangle}/{\langle\mathcal{O}^{\psi}_n\rangle}$, a short
calculation determines it as 1.29 for color-singlet part, and the values of
our fitted ${\langle\mathcal{O}^\psi_8(\bigl.^3\hspace{-1mm}S_1)\rangle}$
and ${\langle\mathcal{O}^\psi_8(\bigl.^1\hspace{-1mm}S_0)\rangle}$ include
the contributions of the feed-down for color-octet part. However, the
feed-down from $\psi^\prime\bigl[{\bigl.^3\hspace{-1mm}P^{(8)}_J}\bigr]$ and
$\chi_{cJ}$ is not considered in our fitting since at NLO it cannot be
properly calculated so far and this omission should be treated as an
approximation.

\begin{figure}[tbp]
\center{
\includegraphics*[scale=0.45]{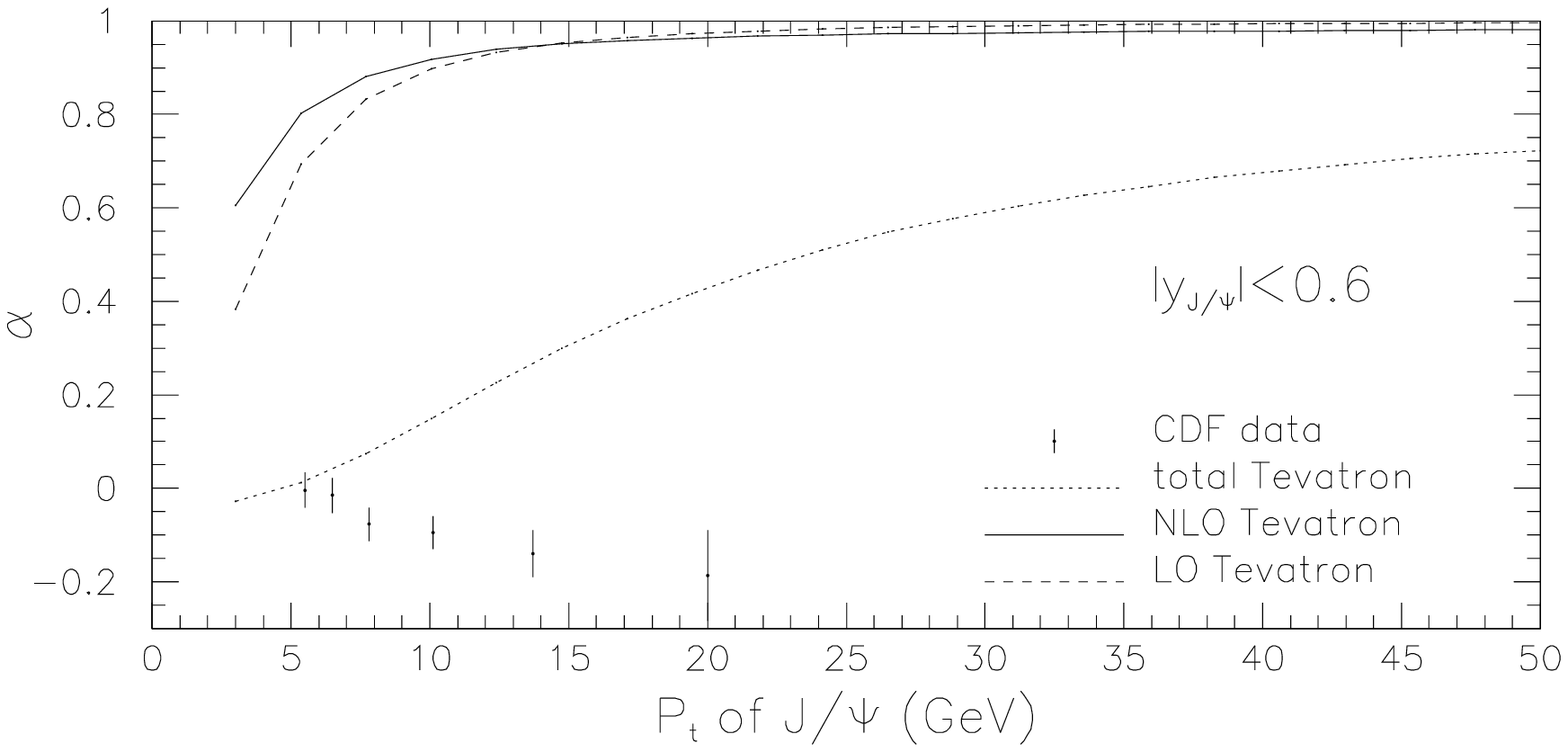}\\
\includegraphics*[scale=0.45]{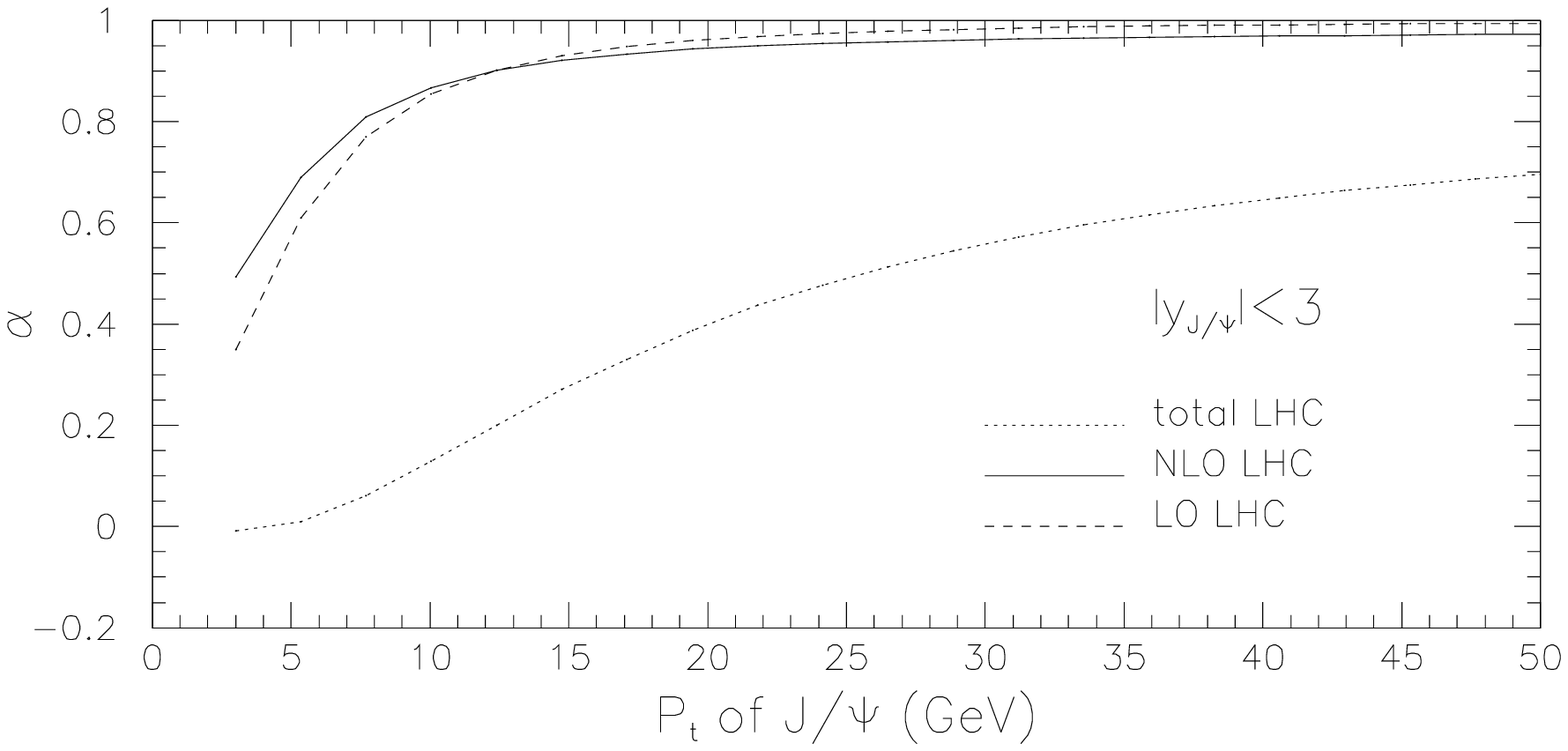}\caption {\label{fig:polar}
Transverse momentum distribution of polarization
$\a$ for prompt $\jpsi$ production, CDF data is from ref~\cite{Abulencia:2007us}.}}
\end{figure}

The dependence of the total cross section on the renormalization
scale $\mu_r $ and factorization scale $\mu_f$ are shown in
Fig.~\ref{fig:total}. It is obvious that the NLO QCD corrections
make such dependence milder. The $p_t$ distributions of ${J/\psi}$
production are presented in Fig.~\ref{fig:pt}
where only slight change appears when the NLO QCD corrections is included. ${%
J/\psi\bigl[^1\hspace{-1mm}S^{(8)}_0\bigr]}$ produces unpolarized ${J/\psi}$
for both LO and NLO. The $p_t$ distributions of ${J/\psi}$ polarization
parameter $\alpha$ for ${J/\psi\bigl[^3\hspace{-1mm}S^{(8)}_1\bigr]}$ are
shown in Fig.~\ref{fig:polar} and there is a slight change when the NLO
corrections are taken into account.

As a summary, in this work, we have calculated the NLO QCD corrections to ${%
J/\psi}$ production via color-octet states ${J/\psi\bigl[\bigl.^1\hspace{-1mm%
}S^{(8)}_0,\bigl.^3\hspace{-1mm}S^{(8)}_1\bigr]}$ at Tevatron and LHC. With $%
\mu_r=\mu_f=\mu_0=\sqrt{(2m_c)^2+p_t^2}$, transverse momentum cut $p_t>3$GeV 
and rapidity cut $|y|<0.6$
(Tevatron) and $|y|<3$(LHC) for $J/\psi$,
the K factors of total cross section (ratio of NLO to
LO) are 1.235 and 1.139 for ${J/\psi\bigl[^1\hspace{-1mm}S^{(8)}_0\bigr]}$
and ${J/\psi\bigl[^3\hspace{-1mm}S^{(8)}_1\bigr]}$ at Tevatron, while at LHC
they are 0.826 and 0.800 respectively. Unlike for the color-singlet case,
there are only slight changes to the transverse momentum distributions of ${%
J/\psi}$ production rate and the ${J/\psi}$ polarization when the NLO QCD
corrections are taken into account. The results imply that the perturbative
QCD expansion quickly converges for ${J/\psi}$ production via the S-wave
color-octet state, in contrast with that via color-singlet, where the NLO
contributions are too large to hint a convergence at the NNLO.
As shown in Fig.~\ref{fig:polar}, an obvious gap  between the
theoretical results for $J/\psi$ polarization calculated up to NLO
and the experimental measurements at Tevatron  is observed, even
though both color singlet and octet are included. In the well
established theoretical framework of NRQCD there still remains a
narrow window which might make up the gap, namely one needs to
investigate the NLO corrections to ${J/\psi}$ production via P-wave
color octet state and ${J/\psi}$ production by feed-down from
$\chi_{cJ}$. It is unclear how the situation would be when
contributions from these two sources at NLO are taken into account,
as we know that NLO QCD corrections to P-wave
state in $e^++e^-\rightarrow J/\psi+\chi_{c0}$ evaluated in Ref.~\cite%
{Zhang:2008gp} are very large. However, a careful analysis indicates
that it is not really the case because as aforementioned, the P-wave
color-octet was not observed at B-factory experiments. Even though
the P-wave color-octets do contribute, in analog to the case for the
S-wave color-octet, it is reasonable to assume that the NLO QCD
correction to the P-wave color-octet is not too large due to the
same power counting of the $p_t$ distribution behavior. Then the
$p_t$ distribution of the P-wave color-octets will be almost the
same as the color-octet $^1S_0$ at NLO. It means that the fitting at
NLO, while including the color-octet $^1S_0$ state, can be thought
as the P-wave color-octet part is also included. Then a definite
conclusion is drawn that the huge discrepancy of ${J/\psi}$
polarization between theoretical predication and the experimental
measurement cannot be solved by just including NLO QCD corrections
within NRQCD framework and one should explore real solution along
other lines.

This work is supported by the National Natural Science Foundation of China
(No.~10475083) and by the Chinese Academy of Sciences under Project No.
KJCX3-SYW-N2.
\bibliographystyle{plain}
\bibliography{paper}

\end{document}